\documentclass[twocolumn,twoside]{revtex4}
\usepackage{graphicx}
\usepackage{fancyhdr}
\usepackage{amsmath}

\usepackage{float}

\newcommand\barparen[1]{\overset{\textbf{\fontsize{1pt}{1pt}\selectfont(---)}}{#1}}

\newcommand\dCP{\delta_{\mathrm{CP}}}
\newcommand\JCP{J_{\mathrm{CP}}}

\newcommand\numu{\nu_{\mu}}
\newcommand\numub{\bar{\nu}_{\mu}}
\newcommand\numup{\barparen{\nu}_{\mu}}

\newcommand\nue{\nu_{e}}
\newcommand\nueb{\bar{\nu}_{e}}
\newcommand\nuep{\barparen{\nu}_{e}}

\newcommand\Dm[1]{ \Delta m^2_{#1} }

\begin{document}

\title{CP-violation search with T2K data}


\author{J.~G.~Walsh\\
on behalf of the T2K Collaboration}
\affiliation{Michigan State University, East Lansing, MI, USA, 48824}

\begin{abstract}
The T2K experiment is a long-baseline neutrino oscillation experiment which uses $\nu_{\mu}$ and $\bar{\nu}_{\mu}$ beams to constrain CP-violating effects in a 3-flavor PMNS neutrino mixing model. Through $\numu\to\nue$ and $\numub\to\nueb$ appearance channels, T2K is sensitive to CP-violating effects in neutrino mixing. An excess of $\nue$ candidates in the $\nu$-beam mode is observed when compared to the CP conserving cases. T2K finds a best fit value of $\dCP=-1.97_{-0.70}^{+0.97}$ using  Feldman-Cousins corrected intervals and excludes CP-conserving values of $\dCP$ of $0$ and $\pi$ at the 90\% CL. $\JCP=0$ is also excluded at 2$\sigma$ when using a flat prior in $\dCP$, favoring negative values.

\end{abstract}

\maketitle

\thispagestyle{fancy}

\section{Introduction}
The Tokai to Kamioka (T2K) experiment~\cite{T2KNIM} is a long baseline neutrino oscillation experiment in Japan which looks for $\barparen{\nu}_{\mu}$ disappearance and $\barparen{\nu}_e$ appearance in an almost pure $\barparen{\nu}_{\mu}$ beam. T2K is able to run in both neutrino and antineutrino mode with an off-axis narrow band beam peaked at 600~MeV and a baseline of 295~km. A suite of near detectors 280~m downstream of the beam production target is used to constrain beam and interaction modelling by sampling the unoscillated beam. By measuring the rate of appearance of $\barparen{\nu}_e$ in the $\barparen{\nu}_{\mu}$ beams, T2K is sensitive to CPV effects in neutrino mixing, and able to constrain the value of the CPV term, $\delta_{\mathrm{CP}}$, in the three-flavor PMNS model~\cite{Pontecorvo}\cite{MNS}. This parameter is degenerate with other mixing parameters, which are constrained by external measurements~\cite{PDG2019} or through the CP-conserving $\nu_{\mu}$-disappearance measurement, presented in later in this conference~\cite{Ali}.
\section{CP-violation in neutrino mixing}
In the three-flavor Pontecorvo-Maki-Nakagawa-Sakata (PMNS) neutrino mixing model, the weak mass eigenstates are related by the $3\times 3$ unitary PMNS matrix, which is represented as the product of three rotation matrices with angles, $\theta_{ij}$ which describe the size of the mixing between mass states $i$ and $j$. An additional complex phase, $\delta_{\mathrm{CP}}$, associated with the $\theta_{13}$ mixing angle, allows for different mixing behaviour in neutrinos and antineutrinos. 
The probability of $\numup\to\nuep$ oscillation, when vacuum like and maximised with respect to $L$ and $E_{\nu}$, is given by

\begin{figure}[h]
\centering
\includegraphics[width=80mm,trim={45 35 45 35},clip]{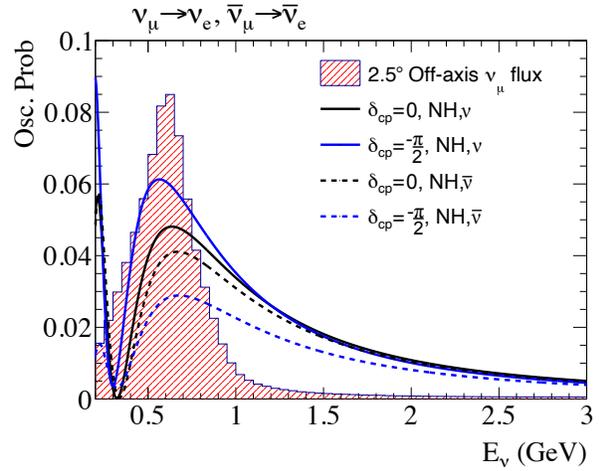}
\caption{Probability of $\numup\to\nuep$ at $L=295~\mathrm{km}$ for different values of the CP-violating phase, $\dCP$. The unoscillated T2K $\numu$ $E_{\nu}$ spectrum is shown in arbitrary units for comparison.} \label{fig:oscprob}
\end{figure}

\small
\begin{align}\begin{split}\label{eqn:NuOscNue}
P ( \nu_{\mu} \rightarrow \nu_{e} ) &\simeq  \sin^2(2\theta_{13})\sin^2(\theta_{23})\sin^2\left(1.27\Delta m^2_{32}\frac{L}{E_{\nu}}\right)\\ &\mp 1.27\Delta m^2_{32}\frac{L}{E_{\nu}} 8J_{\textrm{CP}} \sin^2 \left(1.27\Delta m^2_{32}\frac{L}{E_{\nu}}\right)
\end{split}\end{align}
\normalsize
where the ``Jarlskog invariant'', 
\small
\begin{equation}\label{eqn:NuOscJCP}
J_{\mathrm{CP}}\equiv\sin\theta_{13}\cos^2\theta_{13}
\sin\theta_{12}\cos\theta_{12}
\sin\theta_{23}\cos\theta_{23}
\sin\delta_{\textrm{CP}}
\end{equation}
\normalsize
is a function of $\sin\delta_{\textrm{CP}}$. This gives rise to the opposing suppression and enhancement of $P ( \nu_{\mu} \rightarrow \nu_{e} )$ and $P ( \bar{\nu}_{\mu} \rightarrow \bar{\nu}_{e} )$ respectively, or vice versa if $J_{\mathrm{CP}}$ is non-zero, indicating $\delta_{\textrm{CP}}$ is neither 0 nor $\pi$. A comparison of this effect on the appearance probabilities and the T2K $E_{\nu}$ spectrum is shown in Fig.~\ref{fig:oscprob}.

An eightfold degeneracy exists between $\dCP$, the sign of the mass squared splitting $\Dm{32}$, and the octant of $\theta_{23}$. Mass ordering dependent matter effects can be exploited by increasing baselines to improve the resolution of $\Dm{32}$, but shorter baselines give a purer $\dCP$ sensitivity.

\section{The T2K experiment}
In the T2K experiment~\cite{T2KNIM}, 30~GeV protons from the J-PARC main ring accelerator are impinged on a graphite target producing hadrons. Magnetic horns select charged pions which are then allowed to decay to produce $\numup$. The polarity of the horns selects for positive or negative pions to produce a $\numu$ or $\nue$ beam respectively.
Kaons, wrong-sign pions and muons may undergo decays which contribute wrong-sign and intrinsic $\nuep$ backgrounds to the beam. The beam is then sampled by a suite of near detectors 280~m from the target, and the Super-Kamiokande (Super-K) detector 295~km away in the Kamioka mountains. One of the near detectors, ND280, and the far detector, Super-K, lie on a path which is 2.5$^\circ$ angle to the beam. Another near detector, INGRID, sits on axis and measures the beam intensity and direction. This 2.5$^\circ$ off-axis angle creates a beam with a more narrow band of energies which is tuned to maximise oscillation probabilities for T2K's baseline with a peak energy of 600~MeV. This also reduces the intrinsic $\nue$ background due to the different parent kinematics. 

Super-K is a 50~kton water Cherenkov detector, which can distinguish between $\numu$ and $\nue$ events by the shape of the ring left on PMTs which instrument the detector~\cite{SKNIM}. The 600~MeV peak energy is lower than similar experiments, but ensures a large fraction of the events in the detector occur via the charged current quasielastic (CCQE) channel, from which the neutrino energy can be estimated with the least bias in a Cherenkov detector. The off-axis near detector, ND820, is a magnetized tracking detector and constrains the uncertainties in the interaction modelling as well as the intrinsic ``wrong sign'' contamination, i.e. $\bar{\nu}$s in a $\nu$-beam or $\nu$s in a $\bar{\nu}$-beam, which Super-K is unable to do as it is unmagnetized.
\section{The T2K Likelihood}
The T2K likelihood consists of four main model components: the beam; the interactions; the detectors; and the oscillation physics. The first two of these are common to both ND280 and Super-K, each detector response has its own modelling, and then the oscillations are only relevant to the modelling of the far detector event rates. The likelihood of each detector can be fit either sequentially, or simultaneously.

The data at each detector is selected as to best constrain the uncertainties in each part of the model. ND280 data is split by beam mode and final state lepton charge to constrain the flux and wrong-sign backgrounds. Then events are split by the number of reconstructed final state pions and hadrons as CC0$\pi$, CC1$\pi$ and CC-Other, creating samples rich in the signal (CCQE), and dominant backgrounds at T2K energies. The events are also split by target material, in samples which are either entirely hydrocarbon target, or a mixture of hydrocarbon and water.

At the far detector, the data is split into four CCQE-like single ring samples, $\numu$-like (1R$\mu$) and $\nue$-like (1R$e$) for each beam mode, as well as an additional $\nue$-like sample in the $\nu$-beam mode in which the resonant production of a pion is inferred from an additional delayed decay election ring (1R$e$1d.e.). This provides a direct constraint on the dominant background and adds additional statistics to the $\nue$ appearance channel.
\section{Extracting the oscillation parameters}

In one of the two T2K analyses presented, the data at ND280 are used to constrain the beam and interaction models in a binned maximum likelihood fit, along with the ND280 detector response modelling. The best-fit point, along with a covariance matrix, of each of the parameters in the beam and interaction model are propagated to the far detector fit. In the fit to Super-K data, the beam and interaction parameters, as well as the Super-K detector parameters are marginalized over before a grid search of the $\Delta\chi^2$ is performed for of each of the PMNS parameters. In these grid searches, the other PMNS parameters are treated as nuisance parameters and marginalized over to produce $\Delta\chi^2$ surfaces in either one or two dimensions. Confidence levels are constructed for each of the parameters, which is done according to the Feldman-Cousins~\cite{FC} method in the case of $\dCP$.

In the other analysis, the data at ND280 and Super-K can be fit simultaneously using a Markov Chain Monte Carlo (MCMC) method which samples the likelihood at random steps to slowly build a posterior probability distribution. Where the sequential fit makes an assumption of Gaussianity in the propagation of the flux and cross-section parameters from the near detector fit, the full shape of the likelihood is able to be explored in the MCMC fit. Credible intervals are then constructed from these posterior distributions.

Model uncertainties are constrained by a mix of internal and external measurements, most significantly for the constraint on $\dCP$ being a Gaussian prior uncertainty on $\theta_{13}$ taken from the PDG average of the reactor neutrino experiments~\cite{PDG2019}.
\section{The T2K Data}
The data collected correspond to $19.7(16.3)\times10^{20}$ protons-on-target (POT). This is a 33\% increase in $\nu$-mode data collected at Super-K since the previous T2K result~\cite{Nature}\cite{Long2018}, yielding an additional 18 $\nue$ candidates and 75 $\numu$ candidates giving total event rates for each of the five samples as shown in table~\ref{tab:dCPEvents}.

\begin{table}[htbp]
\centering
\begin{tabular}{l c|cccc|c}
\hline
\hline
\multicolumn{2}{c|}{Sample} & \multicolumn{4}{c|}{True $\dCP$ (rad.)} & Data\\
\multicolumn{2}{c|}{}           & $-\pi/2 $ & $0$ & $ \pi/2$ & $  \pi$ & \\
\hline
1R$\mu$ & $\nu$-mode  & $346.61$  & $345.90$  & $346.57$  & $347.38$  & $318$     \\
                      & $\bar{\nu}$-mode  & $135.80$  & $135.45$  & $135.81$  & $136.19$  & $137$     \\
\hline
1R$e$  & $\nu$-mode  & $96.55$   & $81.59$   & $66.89$   & $81.85$   & $94$      \\
                      & $\bar{\nu}$-mode  & $16.56$   & $18.81$   & $20.75$   & $18.49$   & $16$      \\
\hline
1R$e$1d.e.                 & $\nu$-mode  & $9.30$    & $8.10$    & $6.59$    & $7.79$    & $14$      \\
%\hline
%\multirow{2}{*}{\rmu $E_{rec}<1.2~\text{GeV}$} & $\nu$-mode & $209.14$ & $208.80$ & $209.11$ & $209.57$ & $191$ \\
                                              % & $\bar{\nu}$-mode & $68.09$ & $67.90$ & $68.09$ & $68.30$ & $71$ \\
\hline
\hline
\end{tabular}
\caption{Event rate prediction for $\numu$ and $\nue$ candidates like using oscillation parameters and systematic parameters at best-fit while varying $\dCP$.}
\label{tab:dCPEvents}
\end{table}

\begin{figure}[h]
\centering
\includegraphics[width=75mm, trim={8 5 17 13},clip]{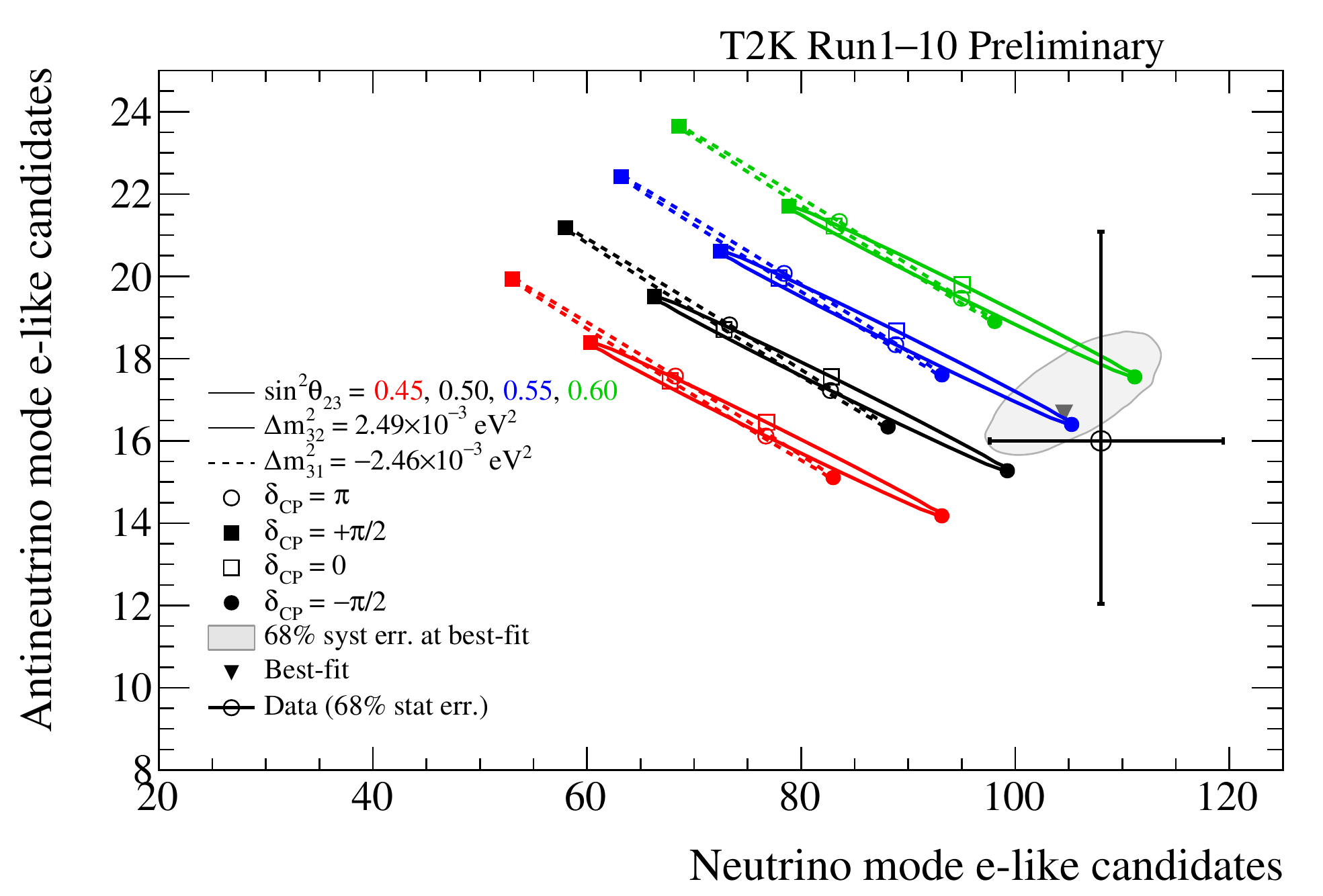}
\caption{Predictions of the number of $\nue$-like candidates in each beam mode for different values of the PMNS parameters compared to the data observed at T2K. Alignment of maximally CP-violating values of $\dCP$ across each mass ordering due to vacuum-like oscillations gives T2K better sensitivity in the region in which its data lie.} \label{fig:biprob}
\end{figure}

The $\nue$ and $\nueb$ event rates are compared to the different predictions of the PMNS model indicated points on the the different ellipses in Fig.~\ref{fig:biprob}. An excess in the $\nu$-mode $\nue$-like event rate compared with the CP-conserving cases can be seen, particularly in the CC1$\pi$-like sample. This excess has decreased upon the collection of more data since the previous result~\cite{Nature}. It is less clear what value of $\dCP$ is preferred in the $\nueb$-like data, due to the smaller $\bar{\nu}$ cross section in matter yielding a lower event rate.

\section{Constraints on $\dCP$ and $\JCP$}
\begin{figure}[h]
\centering
\includegraphics[width=80mm,trim={0 2 0 10mm},clip]{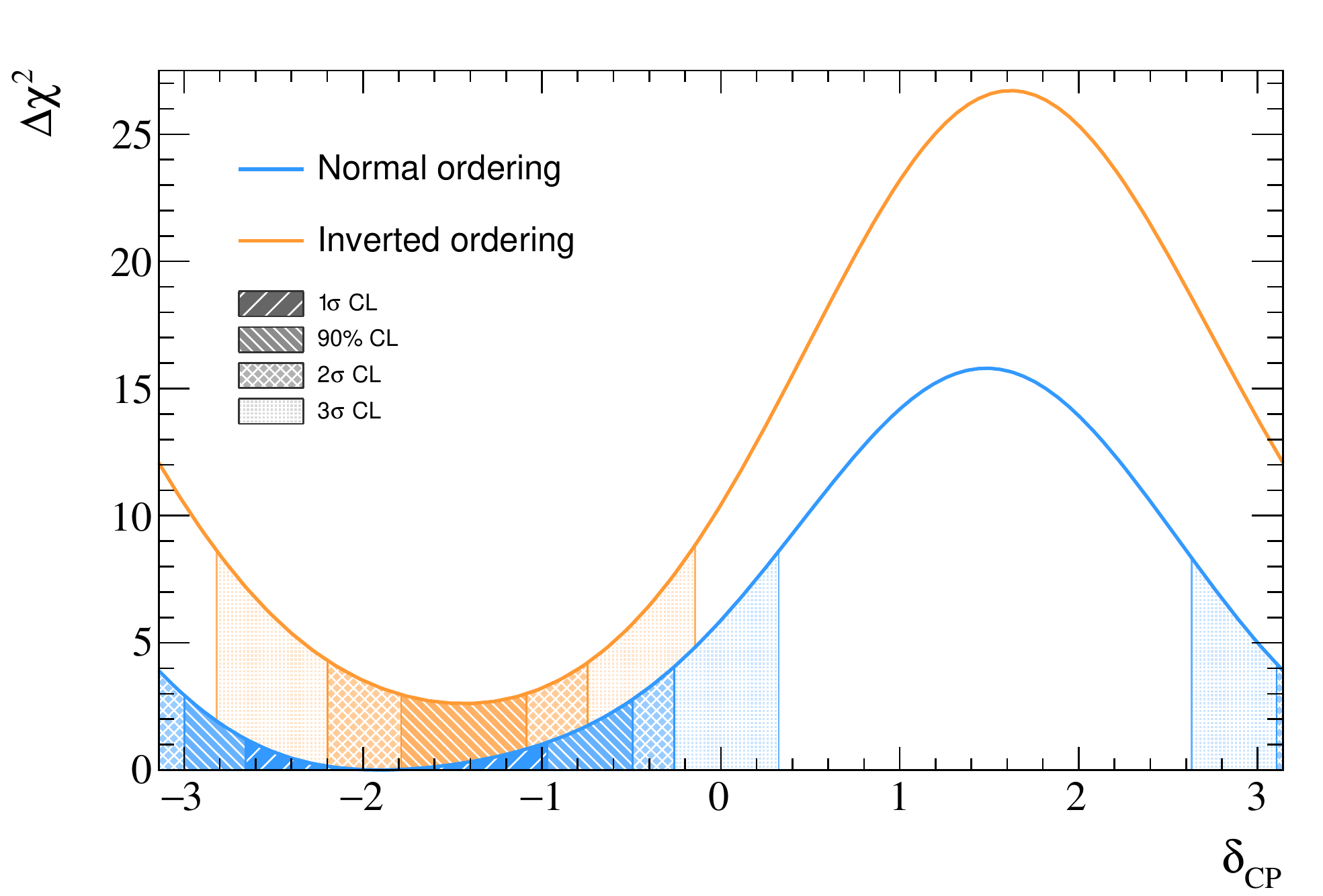}
\caption{$\Delta\chi^2$ surface for $\dCP$ and with Feldman-Cousins confidence level intervals for both mass orderings. A large region around $\pi/2$ is excluded to 3$\sigma$ or more in both mass orderings. CP conserving values are excluded at the 90\% confidence level.} \label{fig:ptheta_dcp}
\end{figure}

Presented here are T2K's constraints on $\dCP$ and $\JCP$, the constraint on $\theta_{23}$ and $\Dm{132}$ from the $\numup$ disappearance channel are presented in more detail in~\cite{Ali}. T2K excludes 35\% of values of $\dCP$ around $+\pi/2$ to 3$\sigma$ or greater, preferring values closer to $-\pi/2$, as seen in Fig.~\ref{fig:ptheta_dcp}. The CP-conserving values, 0 and $\pi$ are excluded to at the 90\% confidence level and from the 90\% credible interval under normal ordering. A larger range of values are excluded in inverted ordering, but the preference is slight and normal and inverted ordered best-fit values are consistent with each other. The best-fit value of $\dCP$ is at -1.97 and is less maximal than T2K's previous result of -1.74, which corresponds to the reduction in the excess beyond the PMNS-like predictions. 
This is statistically dominated and the exclusion statements are robust when the impact of missing systematic effects from alternative interaction models are tested.

% \begin{figure}[h]
% \centering
% \includegraphics[width=80mm]{Figures/MaCh3_deltaCP.pdf}
% \caption{Posterior probability distribution for $\dCP$ marginalized over both mass orderings with credible intervals. A large region around $\pi/2$ is excluded to 3$\sigma$ or more and CP conserving values are excluded at the 90\% level.} \label{fig:mach3_dcp}
% \end{figure}

\begin{figure}[h]
\centering
\includegraphics[width=65mm,trim={0 9 0 16mm},clip]{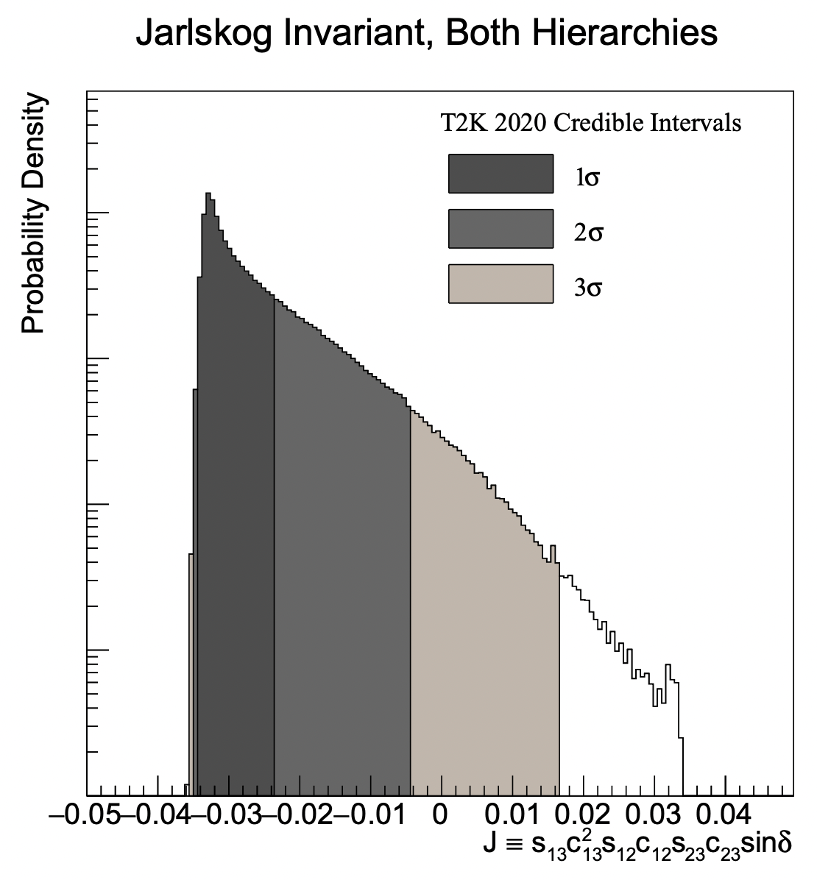}
\caption{Posterior probability density for the Jarlskog invariant $\JCP$ with a flat prior in $\dCP$ marginalized over both mass orderings.} \label{fig:jarlskog}
\end{figure}

The $\JCP$ posterior density and credible intervals are shown in Fig.~\ref{fig:jarlskog}. The T2K data is more consistent with negative values of $\JCP$, excluding CP conservation at 2$\sigma$ when the prior on $\dCP$ is flat. CP conservation is excluded at 90\% when a flat prior on $\sin\dCP$ is used. The $\JCP$ posterior and its credible intervals are shown in Fig.~\ref{fig:jarlskog}

\section{Potential of joint fits with other experiments}

\onecolumngrid
\begin{figure*}[tbp!]
\centering
\includegraphics[width=140mm,trim={10 1 10 15},clip]{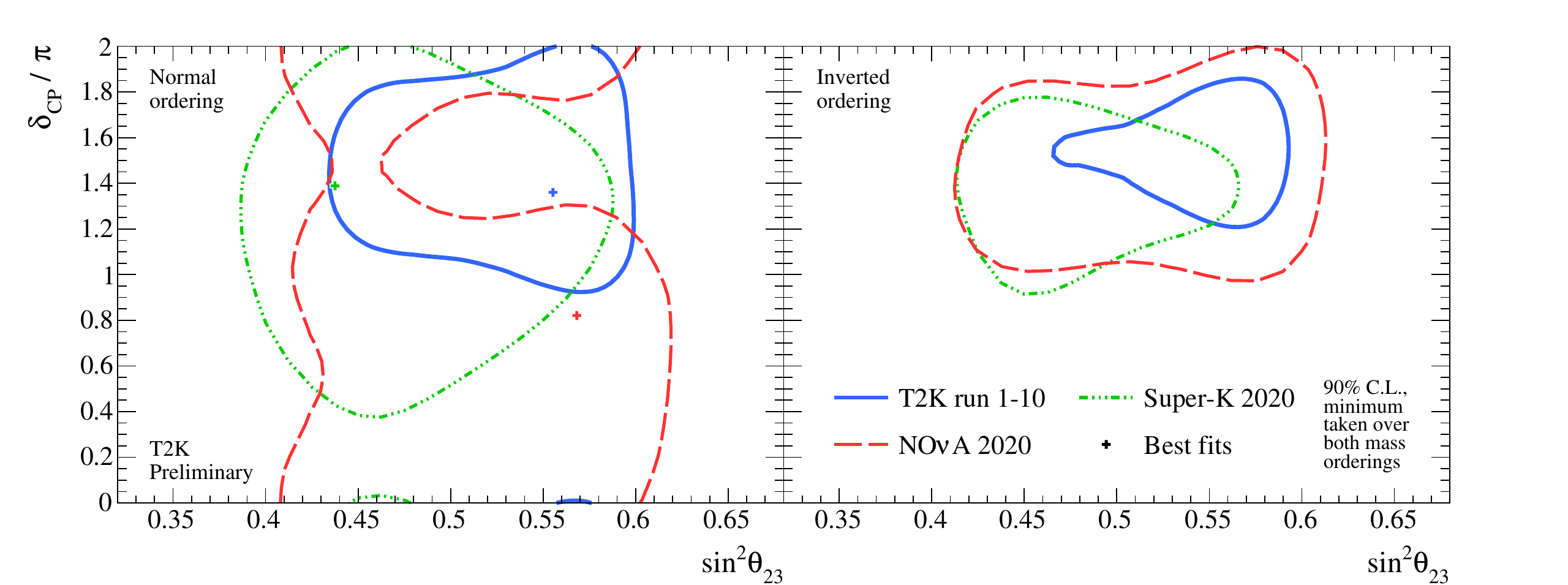}
\caption{Comparison of the T2K constraint on $\dCP-\theta_{23}$ to the NOvA and Super-K experiments. T2K and Super-K have very consistent best fit values of $\dCP$ but prefer different octants of $\theta_{23}$. T2K and NOvA both prefer the upper octant of $\theta_{23}$, but have best fit points which sit outside of each other's 90\% confidence level.} \label{fig:t2ksknova}
\end{figure*}
\twocolumngrid

A comparison of the T2K $\dCP$-$\sin^2\theta_{23}$ constraint to those of Super-K~\cite{SK2020} and NOvA~\cite{NOvA2020} is shown in figure~\ref{fig:t2ksknova}. T2K's best fit value of $\dCP$ is very consistent with that of Super-K but lies outside of NOvA's 90\% confidence limit contours. There does, however, remain overlap at of the 90\% confidence levels across both mass orderings, and while T2K does not strongly prefer a mass ordering, NOvA's constraint has a $\dCP$ dependence to this preference, and its best fit value for $\dCP$ within the inverted ordering is closer to that of T2K's. 

A joint fit of the T2K and NOvA data should break some of the degeneracy between T2K and NOvA by exploiting the more vacuum like measurement of T2K and the stronger mass ordering dependent matter effects of NOvA. Similarly, there are very strong matter effects over the 13,000~km baseline of Super-K, as well as $\theta_{23}$- and $\dCP$-dependent effects on the Super-K spectra which T2K can constrain.

In addition to these joint analyses with other experiments, upgrades to the T2K experiment~\cite{NDUPtdr} will allow for better constraints on the PMNS parameters. These are discussed in more detail in ~\cite{Ali}.

\section{Summary}
T2K has the current world leading constraint on the value of $\dCP$ and excludes CP-conservation to the 90\% confidence level and from its 90\% credible intervals. Though there are differences between the T2K and NOvA best fit values, the 90\% confidence intervals have overlap, particularly in the case of inverted ordered neutrino masses. A joint fit is underway to resolve this difference and will exploit the different sensitivities of each experiment to break apart the highly degenerate PMNS mixing parameter space. Similar fits are underway between T2K and Super-K which has a very large baseline for its upward going atmospheric neutrinos.

\bigskip % extra skip inserted

\end{document}